\documentclass[pra, final, showpacs, twocolumn, groupedaddress, amsfonts]{revtex4}

\usepackage{amsmath, isolatin1, graphicx, color, amsxtra, amssymb, bm}

\newcommand{\beqn}{\begin{eqnarray}}
\newcommand{\eeqn}{\end{eqnarray}}
\newcommand{\refeqn}[1]{Eq.~(\ref{#1})}
\newcommand{\reffig}[1]{Fig.~\ref{#1}}

\newcommand{\ts}{\text{s}} 
\newcommand{\ti}{\text{i}} 
\newcommand{\tp}{\text{p}} 
\newcommand{\tc}{\text{c}}
\newcommand{\td}{\text{d}}

\newcommand{\tacc}{\text{acc}}
\newcommand{\tcor}{\text{cor}}

\begin{document}

\title{Characterization of an asynchronous source of heralded\\
  single photons generated at a wavelength of 1550 nm}

\author{Maria Tengner} \email[Corresponding author. \!Electronic
address:\! ]{mariate@imit.kth.se} \homepage{http://www.ict.kth.se/MAP/QEO}
\author{Daniel Ljunggren} 
\affiliation{Department of Microelectronics and
  Applied Physics, Royal Institute of Technology, KTH,
  Electrum 229, SE-164 40 Kista, Sweden}

\date{\today}

\begin{abstract}
  
  We make a thorough analysis of heralded single photon sources 
  regarding how factors such as the detector gate-period, the photon 
  rates, the fiber coupling efficiencies, and the system losses 
  affect the performance of the source. In the course of this we 
  give a detailed description of how to determine fiber coupling 
  efficiencies from experimentally measurable quantities. We show 
  that asynchronous sources perform, under most conditions, better 
  than synchronous sources with respect to multiphoton events, but 
  only for nearly perfect coupling efficiencies. We apply the theory 
  to an asynchronous source of heralded single photons based on 
  spontaneous parametric downconversion in a periodically poled,
  bulk, KTiOPO$_4$ crystal. The source generates light with highly
  non-degenerate wavelengths of $810$ nm and $1550$ nm, where the
  $810$ nm photons are used to announce the presence of the $1550$ nm
  photons inside a single-mode optical fiber. For our setup we find 
  the probability of having a $1550$~nm photon present in the single-mode 
  fiber, as announced by the $810$ nm photon, to be $48\%$. The probability 
  of multiphoton events is strongly suppressed compared to a Poissonian
  light source, giving highly sub-Poisson photon statistics. 

\end{abstract}


\maketitle

\section{Introduction}
Sources of single photons are fundamental building blocks in all
areas of quantum information processing using photonic qubits,
such as linear-optics quantum computing \cite{KLM01} and quantum communication.
Consequently, many types of single photon sources have been developed,
e.g.~molecule or atom emission \cite{DDM96, LM00}, nitrogen vacancies
in diamond \cite{KMZW00,BBGVPG02}, and quantum dots \cite{PSVZSPY02, 
ZJBJPSKKEB02}, all having different properties like repetition rate, 
single-photon probability, and emission frequency. A promising
alternative is so-called heralded single-photon sources (HSPS)
\cite{PJF04,FATBBGZ04,TOS04,AOBT05,USBBW05,MAKW02}, where photon 
pairs produced by spontaneous parametric downconversion (SPDC) are 
used to prepare conditional single photons \cite{HM86}. One property
of single-photon sources, essential to most applications, is that the
single photons are prepared in a well defined temporal and spatial mode. 
In contrast to most other sources, HSPS have shown to successfully meet 
the spatial mode requirement by optimizing the coupling into single-mode 
fibers \cite{KOW01a,LT05,CDSM05}. However, there is still room for 
improvements on the photon statistics in time, here referred to as 
the temporal mode. Moreover, HSPS via SPDC also provide a great 
flexibility in the choice of frequency for the single photons.

The basic idea of HSPS can be simply stated as having the detection
event of one of the single-photons of a pair announce the presence
of its partner.  The name ``heralded'' originates from the fact that
the single-photons are not created on demand but rather announced by
an external electrical signal. In the realization presented here, this
signal is asynchronous due to the use of a continuous wave (CW)
pump laser for the SPDC process, in contrast to when a synchronously 
pulsed pump laser is used \cite{PJF04,USBW04,CDSM05}. For HSPS of 
both sorts one can avoid ``empty pulses'' to a high degree, in contrast 
to when weak coherent pulses are used as single-photon sources. In 
essence, the temporal statistics of the heralded photons is controlled by 
utilizing {\it a priori} information extracted from the photon pairs. 
For pulsed sources, as long as the coherence time of the emission, 
$\Delta t_{\tc}$, is longer than the duration of the pump pulse (easily 
obtained when using ultrashort (fs) pulsed pump lasers), a single 
process of stimulated emission will take place \cite{RSMATZG04} giving an 
original photon number distribution (i.e.~the distribution before the 
heralding) which is thermal \cite{MG67}. In contrast, when a CW pump 
laser is used, as long as $\Delta t_{\tc}$ is much shorter than the 
gate-period of the detector, a large number of mutually incoherent SPDC 
processes will be present, each thermally distributed in photon number, 
but collectively giving a Poisson distribution \cite{RSMATZG04,MSNI05}. 
Hence, we have different original distributions which can then be 
altered by the heralding. This gives the opportunity to choose between 
the two cases depending on the needs and requirements of a specific 
application. The Poisson distribution obtained with a CW pump is 
suitable for single-qubit applications like, e.g., quantum cryptography 
where it is essential for security to have few multiphoton events, while 
a single thermal distribution is needed in multiqubit applications, 
where different qubits need to be in the same temporal single-mode for 
interference-effects to take place. The latter property is e.g.~important 
for the realization of logic gates for qubits. 

In this paper, we analyze heralded single photon sources, giving an 
experimental method to characterize HSPS in order to determine 
photon rates and fiber coupling efficiencies, with the goal to fill 
the empty space between theory and experiments. 
We describe in detail how to determine relevant fiber coupling efficiencies 
and photon rates from experimentally measured quantities such as detected 
photon rates, detector efficiencies, dark counts etc., factors all affecting 
the performance of the source. We give a straight-forward scheme to 
determine coupling efficiencies from experimental data not only for HSPS, 
but for other fiber-coupled downconversion sources as well. 
We compare the use of a CW pump and a pulsed pump in HSPS. 
The temporal selection 
made by conditional gating applied to one of the photons in a photon 
pair emitted from a continuously pumped SPDC process modifies the 
photon number statistics. By determining the autocorrelation value 
$g^{(2)}(0)$ of the heralded photon from the coupling efficiencies 
and photon rates we show that either super-Poissonian, 
Poissonian, or sub-Poissonian behavior can be obtained depending on 
the chosen gate-period of the detector and the heralding rate. In 
addition to lowering the probability of empty gates (corresponding to pulses), 
the probability for more than one photon occupying a gate, can now 
also be decreased by using a shorter gate-period. 

Following this analysis, we report the experimental results of a source 
of heralded single-photons created by a quasi-phase-matched nonlinear 
crystal made of periodically poled potassium titanyl phosphate (KTiOPO$_4$). 
The heralded photons have a wavelength of 1550 nm, which makes them suitable 
for transmission in an optical telecommunication fiber, and the heralding 
photons have a wavelength of 810 nm, suitable for efficient detection. 
To characterize the source we use the second-order autocorrelation function, 
which we derive formulas for in terms of singles rates, coincidence rates, 
and coupling parameters, assuming that the original photon distribution is
Poissonian. In this way we are able to determine the autocorrelation
function at zero time-delay without needing to perform a Hanbury-Brown
and Twiss correlation experiment \cite{HBT56b}, which is not a
straightforward task for a heralded and gated source \cite{FATBBGZ04}.

The paper is organized as follows. In Sec.~\ref{Sec:theory} we take a
theoretical viewpoint and investigate the prospects for generating
heralded single photons using the photon-pairs created by a CW laser
in a nonlinear crystal. In Sec.~\ref{Sec:coupling_and_rates} we
describe the principal setup of the source and define the coupling
parameters. We also show how these parameters are determined from the
detected and derived photon rates. Section \ref{Sec:probabilities}
discusses the autocorrelation function and other measures to quantify
the source in terms of system parameters.  The result of the
experiment is presented in Sec.~\ref{Sec:experiment}, and we round off
with some conclusions and discussion in Sec.~\ref{Sec:conclusions}.

\section{\label{Sec:theory}Theory}
\begin{figure}[tb]
  \begin{center}
    \includegraphics[scale = 1]{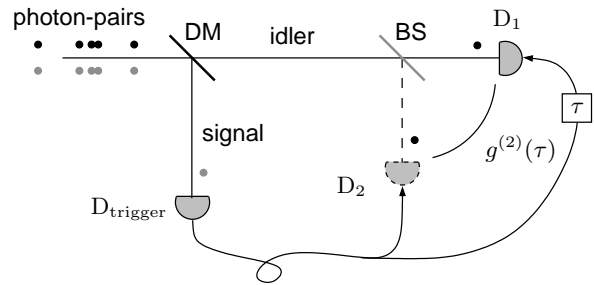}
    \caption{Outline of a heralded single photon source. The
      autocorrelation function $g^{(2)}(\tau)$ can be measured using a
      Hanbury-Brown and Twiss detection scheme using two detectors, or
      $g^{(2)}(0)$ can be measured by a single detector when assuming
      a Poisson distribution in photon number. DM: dichroic mirror; BS:
      beamsplitter.}
    \label{Fig:heralded_princip}
  \end{center}
\end{figure}

The basic principle of the source is depicted in \reffig{Fig:heralded_princip}. 
Using different wavelengths of the trigger photon and the heralded photon 
the two are separated by a dichroic mirror. The trigger photon (signal) 
hits a detector (D$_{\text{trigger}}$) and sends a signal to gate the 
detector (D$_1$) for the heralded photon (idler).
Even for an ideal system, there will be a finite probability for more
than a single photon to arrive within the gate-period of the
detector---a behavior which can be characterized by the second-order
autocorrelation function $g^{(2)}(t_1, t_2)$. (In this section we
assume perfect detectors). The function can be found by a
Hanbury-Brown and Twiss experiment \cite{HBT56b} measuring the
second-order cross-correlation function using two detectors 
(D$_1$ and D$_2$) behind a beamsplitter, see \reffig{Fig:heralded_princip}. 
The true and continuous autocorrelation function is found in the limit 
of infinitely short detector gate-periods, 
${\Delta t_\text{gate} \to 0}$, for different time-delays $\tau = t_1 - t_2$, 
assuming a wide sense stationary and ergodic source of light. In terms 
of probabilities of photon counts, the autocorrelation function is given by
\begin{align}
  g^{(2)}(\tau) = \frac{2P_{m\geq2}(\tau)}{P_{m\geq1}^2(\tau)}.
  \label{Eqn:g2probabilitytau}
\end{align}
where $P_{m\geq k}$ is the probability to find $k$ or more photons
within the detector gate-period. The factor 2 in
\refeqn{Eqn:g2probabilitytau} origins from the fact that the
probabilities are normalized to attain the maximum value of unity,
which is not the case when $g^{(2)}(\tau)$ is written in the standard form
using the intensity of the light. 

Using a single detector, $D_1$, it is clear that as $\tau \to 0$, the
probability for a photon in the idler will be large conditioned on a
photon in the signal, and that the probability of an empty gate is
very small, or even zero, if the probability that the idler photon
makes it from the source to the detector is unity.  If also the
gate-period, $\Delta t_\text{gate}$, is made short, the probability of
two or more photons within the gate becomes small as a result of the
Poisson distribution in the number of photons arriving. 
Hence, by gating in the temporal mode we hereby sub-select events to
effectively change the original statistics. To quantify, we are 
interested in the autocorrelation function of the idler for ${\tau=0}$,
\begin{align}
  g^{(2)}(0) = \frac{2P_{m\geq2}}{P_{m\geq1}^2}.
  \label{Eqn:g2probabilityzero}
\end{align}
It is a well known fact that for ${g^{(2)}(0) < 1}$ and ${g^{(2)}(\tau
  \neq 0) > g^{(2)}(0)}$ we have antibunching, hence sub-Poisson
statistics, and for ${g^{(2)}(0) > 1}$ and ${g^{(2)}(\tau \neq 0) <
  g^{(2)}(0)}$ we have bunching, hence super-Poisson statistics.

We would like to characterize our source using this quantity, which is
zero for perfect antibunching. Thus, we need to know the probabilities
$P_{m\geq2}$ and $P_{m\geq1}$, which can be determined by assuming
that the original distribution is Poisson (a valid assumption as long
as $\Delta t_{\tc} \ll \Delta t_{\text{gate}}$ as will be discussed later), 
and by measuring the mean rate of accidental photons per gate-period, 
${b = \Delta t_{\text{gate}}\bar{R}}$, where $\bar{R}$ is the singles 
rate of accidental photons in counts per second. The rate of accidental 
photons, $\bar{R}$, is simply the difference between the total rate and 
the rate of truly correlated photons. We make the assumption that on 
time scales longer than the coherence time of the photons they can be 
viewed as being independent, so that the correlated photons and the 
accidental photons obey different photon number distributions. (Please 
note that we use the term ``original photon number distribution'' for 
the distribution before heralding. This ``original distribution'' is 
then altered by the conditional gating resulting in the photon number 
distribution of the HSPS.). 
The probability for at least $k$ 
photons to be present in the gate is given by
\begin{align}
  P_{m \geq k} = P^\tcor P_{n \geq k-1}^\tacc + (1 - P^\tcor) P_{n \geq k}^\tacc
  \label{Eqn:Pgk}
\end{align}
where $P^\tcor$ is the probability that the ``true'' twin photon is
present, and $P_{n \geq k}^\tacc$ is the probability that at least $k$
accidental photons are present. The former probability is unity for a
perfect system, and the latter probability is given by
\begin{align}
  P_{n \geq k}^\tacc = 1 - \sum \limits_{j=0}^{k-1} \frac{e^{-b}
    b^j}{j!},
  \label{Eqn:Pagk}
\end{align} 
originating from the original Poisson distribution. Note that 
in \refeqn{Eqn:g2probabilityzero} we do not care if we herald a truly
correlated pair or an accidental, which can happen for lower than
unity coupling efficiencies and transmission factors into the fibers.
\begin{figure}[tb]
    \begin{center}
    \includegraphics[scale = 0.43]{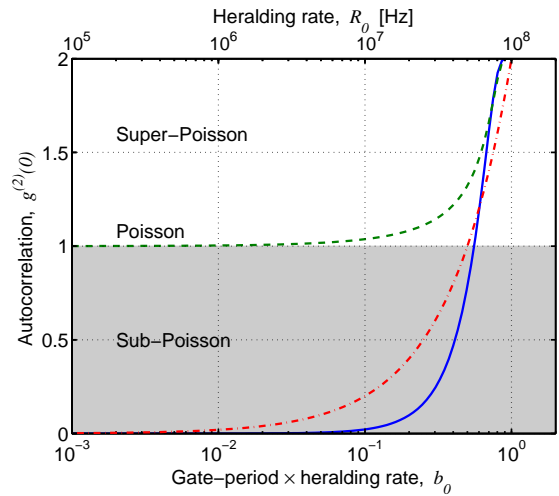}
    \caption{(Color online) The value of the second-order
      autocorrelation function $g^2(\tau)$ at $\tau = 0$, 
      as a function of the parameter $b_0 = \Delta t_\text{gate} R_0$ 
      (bottom $x$-axis), and of the heralding rate $R_0$ (top $x$-axis) 
      for $P^\tcor = 1$, showing either sub- or super-Poissonian photon 
      number statistics. The solid line (blue) shows $g^{(2)}(0)$ for a CW 
      pump plotted against both $b_0$ and $R_0$ (with $\Delta t_\text{gate}=10$ ns). 
      The dashed line (green) represents the statistics
      achieved for a Poissonian source gated at random. The
      dash-dotted line (red) shows $g^{(2)}(0)$ for a pulsed source 
      with pulse rate 100~MHz ($1/\Delta t_\text{gate}$) plotted against 
      the heralding rate $R_0$ (top $x$-axis).}
    \label{Fig:autocorrelation_heralded}
    \end{center}
\end{figure} 
In \reffig{Fig:autocorrelation_heralded} we have plotted
\refeqn{Eqn:g2probabilityzero} for different values of the parameter 
\mbox{$b_0=\Delta t_{\text{gate}} R_0$} (bottom $x$-axis), where 
$R_0$ is the heralding rate. The parameter $b_0$ is related to $b$ 
of \refeqn{Eqn:Pagk} via 
\mbox{$b=\Delta t_{\text{gate}}\bar{R}=\Delta t_{\text{gate}}(R-P^{\tcor}R_0)=\frac{b_0}{1-b_0}-P^{\tcor}b_0$}, 
where the total rate $R$ is assumed to be the same for both signal and idler. 
It is clear from the graph that the statistics of the heralded photons 
can be either sub- or super-Poissonian. The statistics is Poissonian 
for an intermediate value ${b_0 = 0.55}$ for ${P^\tcor = 1}$, and 
${b_0 = 0.42}$ for ${P^\tcor = 0.5}$, given as two examples. 
Sufficiently large values of $b_0$ will always give ``bunched'' 
light in the sense that there will always be more than one photon 
present within the gate-period.  For two uncorrelated events
that are each Poisson distributed, the $g^{(2)}(0)$ value follows
instead the dashed line implying that such a source remains 
Poissonian for short gate-periods or low photon flux, as opposed to 
a HSPS. The expression for $g^{(2)}(0)$ for a CW pumped HSPS
with $P^\tcor = 1$ becomes
\begin{align}
  g^{(2)}(0) = 2[1 - e^{-b}].
  \label{Eqn:g2approx}
\end{align}
In order to compare the CW and the pulsed case, $g^{(2)}(0)$ is also 
plotted as a function of the heralding rate $R_0$ (top $x$-axis). The 
solid line then shows $g^{(2)}(0)$ for a CW source with a fixed gate-period 
\mbox{$\Delta t_{\text{gate}}=10$ ns}, and the dash-dotted line is for a pulsed 
source with pulse repetition rate of \mbox{$1/\Delta t_{\text{gate}}=100$ MHz}. 
As seen, $g^{(2)}(0)$ is higher for a pulsed source than for a CW source in 
the sub-Poisson region, making it more suitable to use a CW pump than a 
pulsed for HSPS. However, this is for the ideal case of perfect coupling 
efficiencies with ${P^{\tcor}=1}$, but in any real experimental situation 
the two choices are practically equivalent as will be discussed later. It 
should be noted that the plotted result is for a pulsed source with an 
original thermal photon number distribution. The 
original distribution will be thermal 
as long as the coherence time of the emission is longer than the duration 
of a pump pulse, $\Delta t_{\tc} > \Delta t_{\tp}$, since there is then a 
single coherent SPDC process present. This situation is rather easily 
achieved by short-pulsed lasers and narrow bandpass filters for the emission, 
or alternatively, with long downconversion crystals to increase the coherence 
length. If ${\Delta t_{\tc} < \Delta t_{\tp}}$ but ${\Delta t_{\tc} > \Delta
t_{\text{gate}}}$ we still have the same situation, but now with the
gate-period as the limiting factor, selecting photons originating from
a single process. However, this situation is rather unrealistic using
pulsed lasers, since it requires ${\Delta t_{\text{gate}} \ll \Delta
t_{\tp}}$. If instead ${\Delta t_{\tc} \ll \Delta t_{\tp}}$ and 
${\Delta t_{\tc} \ll \Delta t_{\text{gate}}}$, there will be a large 
collection of processes, all individually with a thermal distribution, 
but collectively giving a Poisson distribution. Hence, even for a pulsed 
source it is possible to have a Poisson original distribution, but here 
we only consider the thermal case when discussing pulsed sources. Correspondingly, 
for a CW source a thermal distribution is obtained when the coherence time 
is longer than the gate-period, $\Delta t_{\tc} > \Delta t_{\text{gate}}$, 
since then the photons within a gate originate from a single coherent 
SPDC process. However, here we only consider the case when 
$\Delta t_{\tc} \ll \Delta t_{\text{gate}}$, resulting in a large 
collection of SPDC processes collectively giving a Poisson distribution.

For an ideal single photon source, the overall mean photon
number per gate-period, ${\langle n \rangle = b + P^\tcor}$, equals unity, 
which means that ${b = 0}$ and ${P^\tcor = 1}$, i.e. there are no accidental 
photons present and there is perfect correlation between signal and idler. 
In addition, the variance ${\langle \Delta n^2 \rangle}$ of the mean photon 
number should be zero, as quantified by 
${g^{(2)}(0) = 1 + \frac{\langle \Delta n^2 \rangle - \langle n \rangle}{\langle n \rangle^2}}$, 
which motivates why $g^{(2)}(0)$ is a good qualitative measure of HSPS, 
if related to the parameter $b$.

Moreover, the probability for getting exactly $n$ photons
within the gate can also be expressed by the above probabilities as
\begin{align}
  P(n) = P_{m \geq n} - P_{m \geq n+1}.
  \label{Eqn:Pn}
\end{align}
The probability $P(1)$ equals the parameter $\mu^{\text{her}}$
commonly used to characterize sources of single photons, i.e.\ the
probability that exactly one single photon is heralded (ignoring if its a
twin or an accidental for a non-perfect system).

\section{Coupling efficiencies and photon rates}
\label{Sec:coupling_and_rates}
\begin{figure}[!t]
    \begin{center}
    \includegraphics[scale = 1]{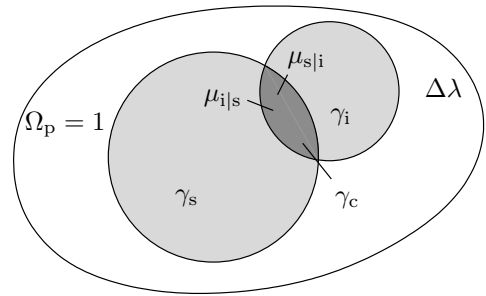}
    \caption{A Venn diagram illustrating the single coupling efficiencies 
    $\gamma_{\ts}$ and $\gamma_{\ti}$, pair coupling $\gamma_{\tc}$, and 
    conditional coincidences $\mu_{\ts\vert\ti}$ and $\mu_{\ti\vert\ts}$. 
    The total amount of pairs within the filter bandwidth $\Delta \lambda$ is 
    denoted $\Omega_{\tp}$ and is normalized to unity.}
    \label{Fig:venndiagram}
    \end{center}
\end{figure}

There are several different coupling efficiencies of interest in
photon-pair sources.  In this section we will define them and discuss
their mutual relations in detail.  For a schematic illustration of the
different quantities see Fig.~\ref{Fig:venndiagram}. All the coupling
efficiencies are related to the bandwidth $\Delta\lambda$ of the
light.  The motivation for this is that the photons emitted from SPDC has in
general a very wide bandwidth, and are preferably filtered before
detection, either by bandpass filters $\Delta\lambda_\text{BP}$ or by
the spectral filtering performed by the single-mode fibers
$\Delta\lambda_\text{SM}$, such that \mbox{$\Delta\lambda \leq
\min(\Delta\lambda_\text{BP}, \Delta\lambda_\text{SM})$}. The
single-mode fiber filtering is an effect of the correlation between each
wavevector's spatial direction and frequency as determined by the
phase-matching in the SPDC process.  By normalizing to the bandwidth
of interest we solely investigate how well photons within that
bandwidth are collected into the fibers.  Hence, as a natural
consequence, with no spatial filtering the ``coupling'' is perfect, as, 
e.g., in the case of a free-space detector or a multimode fiber
(essentially), with a frequency filter in front.

With this in mind, we denote the total number of photon-pairs generated
within a given bandwidth $\Delta\lambda$, with $\Omega_{\tp}$ and
normalize it to 1. This set will of course differ in size in the sense
of absolute numbers of photon pairs, depending on the bandwidth of the
chosen filter.  The {\it single coupling efficiencies} for the signal,
$\gamma_{\ts}$, and idler, $\gamma_{\ti}$, are the fraction of
$\Omega_p$ that is coupled into the single-mode fibers, i.e.~the
probability to have a photon in the fiber which was emitted within the
filter bandwidth $\Delta\lambda$. A high single coupling efficiency
leads to a high photon rate, but does not guarantee a good quality heralded
single-photon source. For that, a high {\it pair coupling efficiency}
$\gamma_{\tc}$, and high {\it conditional coincidences}
$\mu_{\ts\vert\ti}$ and $\mu_{\ti\vert\ts}$ are required.  The pair
coupling efficiency denotes the amount of pairs where both photons are
coupled into the two fibers, i.e.~the degree of overlap 
between the two sets $\gamma_{\ts}$ and $\gamma_{\ti}$ in 
\reffig{Fig:venndiagram}. It is important to note that in general
${\gamma_{\tc} \neq \gamma_{\ts}\gamma_{\ti}}$ , and instead of only
optimizing the single coupling efficiencies it is crucial to maximize
the overlap, i.e.~to couple the matching modes of the signal
and idler into the fibers, in order to obtain a high pair coupling
efficiency \cite{LT05}.  The conditional coincidence is the
probability to have a photon in the fiber given that the partner
photon of the pair is in its fiber.

\begin{figure}[!t]
    \begin{center}
    \includegraphics[scale = 1]{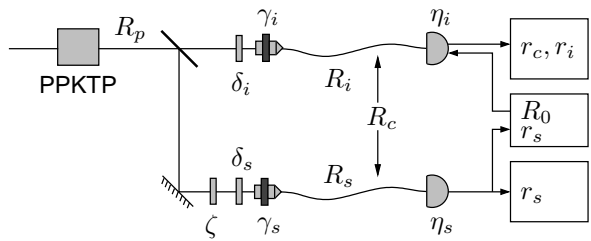}
    \caption{Schematics of the experimental setup showing photon rates and 
      relevant parameters. $R_{\tp}$: rate of generated pairs;
      $\gamma_{\ts}$ and $\gamma_{\ti}$: single coupling efficiencies
      for signal and idler; $\delta_{\ts}$ and $\delta_{\ti}$: total
      transmissions from crystal to detectors; $R_{\ts}$ and
      $R_{\ti}$: total photon rates inside the fibers; $\zeta$:
      compensating factor for unmatched filters between signal and
      idler; $R_{\tc}$: rate of correlated pairs in the fibers;
      $\eta_{\ts}$ and $\eta_{\ti}$: detector efficiencies; $r_{\ts}$:
      detected photon rate for the signal; $R_0$: heralding rate from
      delay generator; $r_{\tc}$: detected heralded rate;
      $r_{\ti}$: detected rate of accidental coincidences.}
    \label{Fig:photonratessetup}
    \end{center}
\end{figure}

All of these coupling efficiencies can be determined from the measured
photon rates and parameters of the experimental setup such as losses
and detector efficiencies. Referring to \reffig{Fig:photonratessetup}, 
we denote by $R_{\tp}$ the total photon pair rate generated within the 
given bandwidth $\Delta\lambda$. The photon rates inside the
single-mode fibers are $R_{\ts}$ and $R_{\ti}$ for the signal and
idler respectively.  They are related to the single coupling
efficiencies by
\begin{equation}
  \gamma_{\ts}=\frac{R_{\ts}}{\zeta\delta_{\ts}R_{\tp}}, \qquad
  \gamma_{\ti}=\frac{R_{\ti}}{\delta_{\ti}R_{\tp}}, 
  \label{Eqn:single_coupling_efficiencies}
\end{equation}
where $\delta_{\ts}$ and $\delta_{\ti}$ are the total transmission
factors for the signal and idler, resulting from the filter transmissions 
and reflection losses of all components between the crystal and the detectors. 
Thus, ${\delta_{\ts}=\delta_{\ti}=1}$ corresponds to an ideal system 
with no losses present other than the fiber
coupling. By weighting the coupling efficiencies by the transmission
factors we obtain measures that solely describe how well the coupling into the
fibers is performed. The factor ${\zeta \leq 1}$ compensates for the
possibly unmatched bandwidths of the interference filters of the
signal and idler. When ${\zeta=1}$ the filter bandwidths match (the
relation between signal and idler for our choice of wavelengths is
${\Delta \lambda_{\ti}\zeta \approx 3.66 \times \Delta \lambda_{\ts}}$)
while ${\zeta<1}$ represents a narrower filter used for the signal than
for the idler.

At the end of the fibers we have single photon detectors with quantum
efficiencies $\eta_{\ts}$ and $\eta_{\ti}$. The signal detector
measure the single photon rate $r_{\ts}$. These detections serve as
the trigger signal to the other detector. However, it is routed via a
delay/pulse generator which in turn provides the gate-pulses for the
idler detector. We call the gate-pulse rate the heralding rate,
denoted $R_0$. This signal announces the presence of the heralded
single photon. In principle $R_0$ should equal $r_{\ts}$, but in
practice $R_0$ is lower because of the dead-time of the
delay/pulse generator used. A heralding pulse gates the idler 
detector for a time $\Delta t_{\text{gate}}$ during which the 
idler photon is expected to arrive at the detector.
From the idler detector we then obtain the measured heralded
photon rate $r_{\tc}$. We also measure the accidental rate
$r_{\ti}$ at the idler detector, i.e.~the single photon rate at random
gating, to provide the mean accidental photon number. 
Also dark count rates, $r_{\ts}^{\td}$ and $r_{\ti}^{\td}$, are measured
for the two detectors, while after-pulsing effects of the \mbox{1550 nm} 
detector are removed by an electrical hold-off circuit (\mbox{10 $\mu$s}).

In order to determine $R_{\tp}$, the photon rate for the
signal is measured using a multimode fiber. This detected rate is 
denoted $r_{\tp}$, and $R_{\tp}$ is then found as
\begin{equation}
  R_{\tp}=\frac{r_{\tp}\alpha_{\tp}^{\text{corr}} -
  r_{\ts}^{\td}}{\eta_{\ts}\zeta\delta_{\ts}}, 
\end{equation}
where $\alpha_{\tp}^{\text{corr}}$ is the correction factor at rate
$r_{\tp}$ for the signal detector, when compensating the detected rate for
the Poissonian distribution of the arrivals of the photons (including the 
dead-time of the detector). The photon rate for the signal inside the 
single-mode fiber, $R_{\ts}$, is obtained in a similar way:
\begin{equation}
  R_{\ts}=\frac{r_{\ts}\alpha_{\ts}^{\text{corr}} -
  r_{\ts}^{\td}}{\eta_{\ts}}. 
\end{equation}
The idler fiber photon rate, $R_{\ti}$, is calculated from the
measured rate of accidental coincidences, $r_{\ti}$, i.e.~the rate when
the idler detector is randomly gated, using 
\mbox{$r_{\ti}=R_0 P_{\text{click}}^{\text{acc}}$}, where
\begin{equation}
  P_{\text{click}}^{\text{acc}}=1 -
  (1-P_{\text{light}})(1-P_{\text{dark}}),  
\end{equation}
is the probability of a detector-click during one gate-period caused by light
or by dark count probabilities.  Assuming a Poisson photon statistics
within the gate, justified by a gate-period $\Delta t_{\text{gate}}$
much larger than the coherence time $\Delta t_{\tc}$ of the
downconverted light, we have \mbox{$P_{\text{light}} = 1 - \exp{(-\eta_\ti
  \Delta t_\text{gate} R_\ti)}$ and $P_{\text{dark}} = \Delta
t_\text{gate} r_{\ti}^{\td}/R_0$}, leading to
\begin{equation}
  R_{\ti}=\frac{1}{\eta_{\ti} \Delta t_{\text{gate}}}\ln \left(
  \frac{1-r_{\ti}^{\td}/R_0}{1-r_{\ti}/R_0} \right).
\end{equation}

The pair coupling efficiency $\gamma_{\tc}$ is defined via the rate of
correlated pairs inside the fibers $R_{\tc}$. This rate describes the
amount of $R_{\tp}$ where both the photons of a pair have coupled into
their respective fiber, giving
\begin{equation}
  \gamma_{\tc} = \frac{R_{\tc}}{\zeta \delta_{\ts} \delta_{\ti} R_{\tp}}.
  \label{Eqn:pair_coupling}
\end{equation}
The correlated pair rate $R_{\tc}$ is determined from the measured
heralded count rate
\mbox{$r_{\tc}=R_0 P_{\text{click}}^{\text{cor}}$}, where 
\begin{equation}
  P_{\text{click}}^{\text{cor}}=1 - (1 -
  P_{\text{light}}^{\tcor})(1-P_{\text{light}}^{\tacc})(1-P_{\text{dark}}),  
\end{equation}
once again is the probability of a detector-click during one gate,
with \mbox{$P_{\text{light}}^{\tcor} = \eta_\ti R_\tc/R_\ts$} as the
probability to detect the ``true'' twin photon, and
\mbox{$P_{\text{light}}^{\tacc} = 1 - \exp{[-\eta_\ti \Delta t_\text{gate}
  (R_{\ti}-R_{\tc}R_0/R_{\ts})]}$} as the probability to detect an
accidental photon. The last minus term in the exponential excludes
those events which are counted as true coincidences.
In terms of photon rates we obtain an implicit expression for
$R_{\tc}$:
\begin{equation}
  \frac{r_{\tc}}{R_0} = 1-\left( 1-
  \eta_{\ti}\frac{R_{\tc}}{R_{\ts}} \right) \left(
  1-\frac{r_{\ti}^{\td}}{R_0} \right) \text{e}^{-\eta_{\ti}
  \Delta t_{\text{gate}} (R_{\ti}-R_{\tc}R_0/R_{\ts})}, 
\end{equation}
which can be solved numerically.

Having determined all the photon rates, we can calculate the different
coupling efficiencies from \refeqn{Eqn:single_coupling_efficiencies},
\refeqn{Eqn:pair_coupling}, and
\begin{equation}
  \mu_{\ti \vert \ts} = \frac{R_{\tc}}{R_{\ts}}, \qquad \mu_{\ts \vert
  \ti} = \frac{R_{\tc}}{R_{\ti}}, 
  \label{Eqn:conditional_coincidences}
\end{equation}
altogether describing how well the fiber coupling is optimized in 
the experiment. Note that $P^{\tcor}$ introduced in Sec.~\ref{Sec:theory} 
equals $\mu_{\ti \vert \ts}$. The conditional coincidences in 
\refeqn{Eqn:conditional_coincidences} are the probabilities of
 having the ``true'' twin photon present, a property which is 
important when using downconversion sources to create entanglement. 
For a HSPS however, the significant quantity is
$\mu^{\text{her}}=P(1)$; the probability to herald exactly one photon,
as determined by \refeqn{Eqn:Pn}.  This procedure to determine rates and
coupling efficiencies is not only relevant for heralded single-photon
sources, but is applicable to other fiber-coupled downconversion
sources as well \cite{LT05, LT05b}.

\section{Heralded single- and multiphoton probabilities}
\label{Sec:probabilities}
As discussed in Sec.~\ref{Sec:theory}, the characterizing quantities
for a heralded single-photon source are the probabilities of the
photon statistics. We will in this section relate these probabilities
to the various photon rates and coupling efficiencies presented in
Sec.~\ref{Sec:coupling_and_rates}.

To obtain the $g^{(2)}(0)$-value for the source we need to determine
the probabilities $P_{m \geq 1}$ and $P_{m \geq 2}$ according to
\refeqn{Eqn:g2probabilityzero}. Using Eqs.~(\ref{Eqn:Pgk}) and (\ref{Eqn:Pagk}), 
expressed in terms of photon rates these probabilities are found to be
\begin{eqnarray}
  P_{m \geq 1} & = & 1-\left( 1-\frac{R_{\tc}}{R_{\ts}} \right)
  \text{e}^{-b}, \label{Eqn:Pm1}\\ 
  P_{m \geq 2} & = & 1-\left[ 1+\left(1-\frac{R_{\tc}}{R_{\ts}}
  \right) b \right] \text{e}^{-b},
  \label{Eqn:Pm2} 
\end{eqnarray}
where $b = \Delta t_{\text{gate}} (R_{\ti}-R_{\tc}R_0/R_{\ts})$.
Inserting this into the expression for $g^{(2)}(0)$,
\refeqn{Eqn:g2probabilityzero}, we obtain
\begin{equation}
  g^{(2)}(0) = \frac{2\left[ 1-\left[ 1+\left(1-\frac{R_{\tc}}{R_{\ts}}
  \right) b \right] \text{e}^{-b}
  \right]}{\left[ 1-\left( 1-\frac{R_{\tc}}{R_{\ts}} \right)
  \text{e}^{-b} \right] ^2}. 
\end{equation}
A good approximation for small $b$ is
\begin{equation}
  g^{(2)}(0) \approx 2(1-\text{e}^{-bR_{\ts}/R_{\tc}}) \approx
  \frac{2bR_{\ts}}{R_{\tc}} 
\end{equation}
for a non-ideal source with $P^{\tcor}=R_{\tc}/R_{\ts}$, in contrast
to \refeqn{Eqn:g2approx}, for which $P^{\tcor}=1$.
Rewriting $g^{(2)}(0)$ using the coupling efficiencies in
\refeqn{Eqn:single_coupling_efficiencies} and
\refeqn{Eqn:pair_coupling} we get
\begin{equation}
  g^{(2)}(0) \approx 2 \Delta t_{\text{gate}} \left(
  \frac{\gamma_{\ts} \gamma_{\ti}}{\gamma_{\tc}}R_{\tp}-R_0
  \right). 
  \label{Eqn:g20approximation2}
\end{equation}
For an ideal antibunched source $g^{(2)}(0)=0$, so we want the value to be as
small as possible. As seen from \refeqn{Eqn:g20approximation2},
$g^{(2)}(0)$ can be made smaller by decreasing the number of
generated photon pairs $R_{\tp}$, i.e.~by simply lowering the pump
power. However, for a single photon source to be useful for
applications, high photon rates are in general desirable, so this does
not seem like a sensible way to improve the performance of the source.
We also conclude that a decrease of the single coupling efficiencies,
$\gamma_{\ts}$ and $\gamma_{\ti}$, and an increase of the pair
coupling, $\gamma_{\tc}$, both lower $g^{(2)}(0)$. Since
$\gamma_{\tc} \leq \min{(\gamma_{\ts},\gamma_{\ti})}$, the optimum is to
have all three equal, but as small as possible. Again however,
this leads to undesirably low photon rates. Decreasing the gate-period
$\Delta t_{\text{gate}}$ is also a possibility, and this seems like a
more natural way to enhance the performance since it essentially does
not affect the photon rates. Yet, $\Delta t_{\text{gate}}$ must still
be kept much longer than the coherence time of the downconverted
photons in order to keep the above analysis valid by maintaining the 
original photon number statistics to be Poissonian.

Using \refeqn{Eqn:Pn}, \refeqn{Eqn:Pm1} and \refeqn{Eqn:Pm2} we find
the expression for $\mu^{\text{her}}=P(1)$ to be
\begin{equation}
  \mu^{\text{her}}=\left( \left( 1-\frac{R_{\tc}}{R_{\ts}} \right) b 
  + \frac{R_{\tc}}{R_{\ts}} \right) \text{e}^{-b}. 
\end{equation}

\section{Experimental results}
\label{Sec:experiment}
\begin{figure}[!t]
  \begin{center}
    \includegraphics[scale = 0.8]{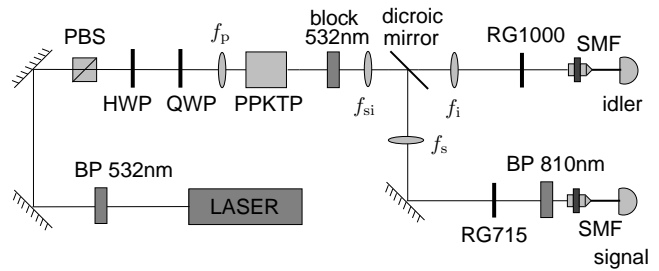}
    \caption{The experimental setup of the heralded single photon
      source.  PBS: polarizing beam splitter; HWP: half wave plate,
      QWP: quarter wave plate; BP: band pass filter; SMF: single-mode
      fiber.}
    \label{Fig:experimentalsetup}
    \end{center}
\end{figure}

The experimental setup of the source is shown in
\reffig{Fig:experimentalsetup}.
A CW laser at a wavelength of 532 nm pumps a 4.5 mm long periodically
poled KTiOPO$_4$ \mbox{(PPKTP)} bulk crystal. The crystal is
poled with a period of 9.6 $\mu$m to assure collinear phase-matching
for a signal and idler at \mbox{810 nm} and \mbox{1550 nm}, respectively. The pump's
polarization is controlled by a polarizing beam splitter, a half wave
plate, and a quarter wave plate, before focusing the light onto the
crystal with an achromatic doublet \mbox{($f_\tp=50$ mm)}. Directly after the
crystal the pump light is blocked by a bandstop filter. The signal and
idler emission are refocused by an achromatic doublet \mbox{($f_{\ts\ti}=30$
mm)} before split by a dichroic mirror, then collimated by two
additional lenses (\mbox{$f_\ts=60$ mm}, \mbox{$f_\ti=40$ mm}), and finally focused
into single-mode fiber by aspherical lenses \mbox{($f=11$ mm)} following
the predictions in \cite{LT05}. In front of the signal fiber-coupler a
Schott-RG715 filter is placed to block any remaining pump light,
together with an interference filter with \mbox{2 nm} bandwidth centered at 
\mbox{810 nm} (all bandwidths are full-width half-maximum, FWHM). For the idler it
suffices with a Schott-RG1000 filter to block the last residue of the pump,
giving an estimated single-mode bandwidth of \mbox{15 nm} for the
accidental photons (set by the spatial filtering of the idler single-mode fiber) 
and \mbox{7 nm} for the coincidence photons (set by the interference filter of 
the signal). The detectors used are a Si-based APD (PerkinElmer SPCM-AQR-14) for the
\mbox{810 nm} light with a quantum efficiency \mbox{$\eta_s=60\%$}, and a homemade
InGaAs-APD (Epitaxx) module for the \mbox{1550 nm} light with \mbox{$\eta_i=18\%$}.
The detection of a \mbox{810 nm} photon triggers the digital delay generator
(DG535 from SRS), which, in turn, generates a gate-pulse for the \mbox{1550 nm} 
detector.

\begin{figure}[tb]
    \begin{center}
    \includegraphics[scale = 0.4]{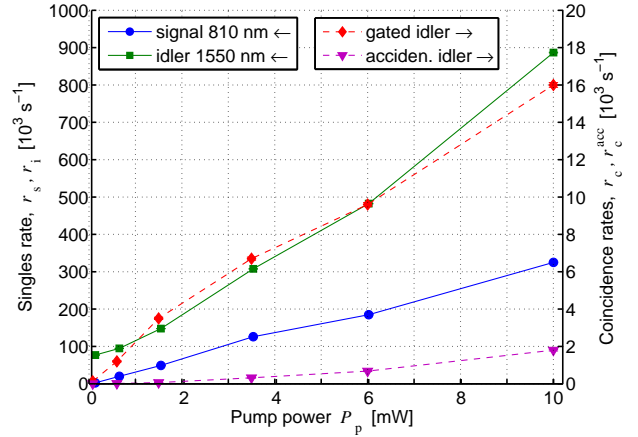}
    \caption{(Color online) The singles rate of signal and idler, both
      in free-running mode (left axis). The idler's rate in counts per
      second is derived from randomly gated mode, with a gate-period
      $\Delta t_\text{gate} = 10\ \text{ns}$, at a rate $R_0$.  The
      right axis shows the total gated heralded rate $r_\tc$ and
      the derived accidental coincidence rate $r_{\tc}^{\tacc}$. 
      Errors are all within the size of the data points in the graph.}
    \label{Fig:ratesvspumppower}
    \end{center}
\end{figure}

We measured the singles- and heralded photon rates for different
pump powers by varying it using neutral density filters. As
expected, both singles, heralded, and accidental counts increase
with the pump power, see \reffig{Fig:ratesvspumppower}. The pump power
1.2 mW was chosen for the subsequent measurements.
\begin{figure}[b]
    \begin{center}
    \includegraphics[scale = 0.37]{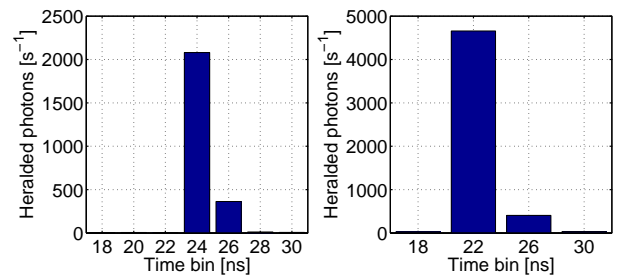}
    \caption{The rate of gated heralded photons, $r_\tc$ for different
      delays of the gate-signal at a heralding rate $R_0 = 65 \times 10^3\ 
      \text{s}^{-1}$. The gate-period, $\Delta t_\text{gate}$, was in
      the left histogram \mbox{2 ns} and in the right \mbox{4 ns}.}
    \label{Fig:jitter_histogram}
    \end{center}
\end{figure}
Histograms of the heralded rate for different delays of the
gate-signal can be seen in \reffig{Fig:jitter_histogram}. 
The gate
delay was moved within a \mbox{12 ns} window for the two cases of
gate-periods, $\Delta t_{\text{gate}}$, of \mbox{2 ns} and \mbox{4 ns}. 
We can observe that the heralded photons are well localized in time
in both cases. The total number of heralded photons are lower for the
\mbox{2 ns} gate-period than for the \mbox{4 ns} gate-period due to
the finite rise time of the gate-pulse, and a lower excess gate
voltage for shorter gate-periods, causing a decrease in the detector
quantum efficiency.
\begin{figure}[t]
  \begin{center}
    \includegraphics[scale = 0.4]{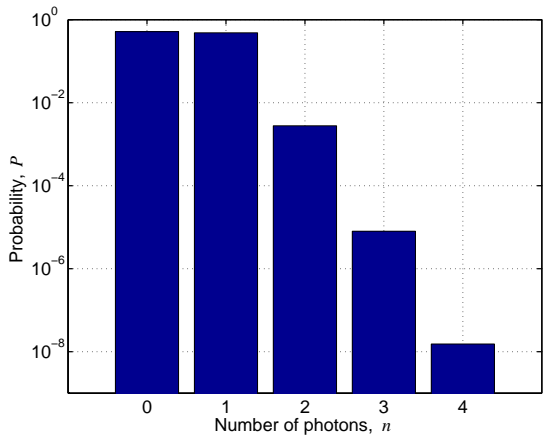}
    \caption{The probability distribution, $P(n)$, of the
      idler photon number, $n$, as a result of gating the idler
      conditioned upon detection of a signal photon. The numbers are
      the results of an experiment at a pump power, $P_\tp = 1.2\ 
      \text{mW}$, $b=0.0057$, 
      and heralding rate $R_0 = 81 \times 10^3\ 
      \text{s}^{-1}$.}
    \label{Fig:heralded_histogram}
    \end{center}
\end{figure}

We have optimized the fiber coupling with the goal of obtaining an as
high conditional coincidence as possible, which did not correspond to
the highest possible single coupling efficiencies. The resulting
detected single counts rate for the signal was $r_{\tp}=218 \times
10^3\ \text{s}^{-1}$ with the multimode fiber, and $r_{\ts}=88 \times
10^3\ \text{s}^{-1}$ with the single-mode fiber. The latter rate
resulted in a heralding rate $R_0=81 \times 10^3\ \text{s}^{-1}$, and a
detected heralded rate $r_{\tc}=7200\ \text{s}^{-1}$ for a
gate-period $\Delta t_{\text{gate}}=10\ \text{ns}$. Accidental
coincidences, i.e.~coincidences measured with random gating, was
$r_{\ti}=130\ \text{s}^{-1}$. The dark count for the signal detector
was $r_{\ts}^{\td}=90\ \text{s}^{-1}$ and for the idler detector
$r_{\ti}^{\td}=40\ \text{s}^{-1}$ at gate-rate $R_0$.
The overall transmission factors in the signal and idler arm were
$\delta_{\ts}=54\%$ and $\delta_{\ti}=63\%$, as determined by sending
strong laser light at the corresponding frequency through the setup
and measuring the loss. The 2 nm interference filter for the signal
and no interference filter for the idler give $\zeta = 0.5$. With
these measured photon rates and setup parameters the actual photon
rates were calculated using the expressions in 
Sec.~\ref{Sec:coupling_and_rates}, obtaining a generated photon-pair rate
$R_{\tp}=1340 \times 10^3\ \text{s}^{-1}$, photon rates inside the
single-mode fibers $R_{\ts}=147 \times 10^3\ \text{s}^{-1}$, and
$R_{\ti}=615 \times 10^3\ \text{s}^{-1}$, and correlated pair rate
inside the fibers $R_{\tc}=71 \times 10^3\ \text{s}^{-1}$.  This
resulted in single coupling efficiencies $\gamma_{\ts}=40\%$ and
$\gamma_{\ti}=71\%$, pair coupling efficiency $\gamma_{\tc}=31\%$, and
conditional coincidences $\mu_{\ti\vert\ts}=48\%$ and
$\mu_{\ts\vert\ti}=12\%$.

\begin{figure}[tb]
  \begin{center}
    \includegraphics[scale = 0.4]{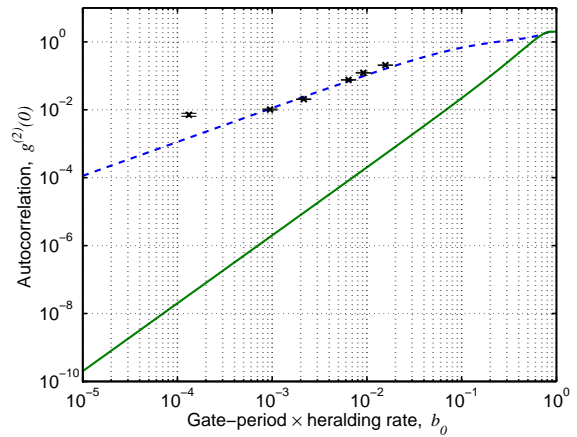}
    \caption{(Color online) The autocorrelation $g^{(2)}(0)$ as a function of 
      \mbox{$b_0=\Delta t_{\text{gate}} R_0$} with $\Delta
      t_{\text{gate}}=10$ ns. The solid line (green) is the theoretical 
      curve with $P^{\tcor}=R_{\tc}/R_{\ts}=1$. The dashed line (blue)
      is the theoretical curve with the experimental 
      coupling efficiencies and ${P^{\tcor} < 1}$. The crosses are the experimental 
      data (with error bars) where $b_0$ has been varied by changing the pump power 
      (0.08, 0.6, 1.5, 3.5, 6, and 10 mW). For the experimental data 
      the heralding rate $R_0$ has been compensated for experimental 
      limitations such as e.g.~detector quantum efficiency in order 
      for a fair comparison with theory.}
    \label{Fig:g2vsb0}
  \end{center}
\end{figure}

With the calculated photon rates the heralded photon statistics was
determined, see \reffig{Fig:heralded_histogram}.
The probability to have zero photons present within the gate-period
was $P(0)=0.514 \pm 0.003$, and the probability to have exactly one photon
present was $\mu^{\text{her}}=P(1)=0.483 \pm 0.003$. The probabilities for
higher number of photons drop off rapidly, with
$P_{m \geq 1}=0.486 \pm 0.003$, and 
$P_{m \geq 2}=0.0028 \pm 0.00002$, giving $g^{(2)}(0)=0.0235 \pm 0.0005$. For the different
pump powers in \reffig{Fig:ratesvspumppower}, $g^{(2)}(0)$ was also
calculated, showing a growth with pump power via the $b_0$
parameter, see \reffig{Fig:g2vsb0}, in agreement with
\refeqn{Eqn:g20approximation2}. We see that $g^{(2)}(0)$ decreases faster 
with decreasing $b_0$ for the ideal case (solid line), where ${P^{\tcor}=1}$, 
than for the non-ideal case. The non-ideal CW case will in fact approach 
the pulsed case (illustrated by the lower dashed line in 
\reffig{Fig:autocorrelation_heralded}), making them practically 
equivalent in a real experimental situation. Indeed, the observation we 
have made is that the more advantageous behavior of the ideal CW case 
(in terms of correlation statistics) is cancelled as soon as the coupling 
efficiencies decrease even slightly below unity, thus 
rapidly reducing the CW case to the pulsed case.

\section{Conclusions and discussion}
\label{Sec:conclusions}
In this paper we have made an analysis of an asynchronous heralded
single-photon source in terms of photon rates, gate-periods, coupling
efficiencies etc. We have determined the photon number statistics and
found it to be highly sub-Poissonian. We have also calculated the
autocorrelation $g^{(2)}(\tau = 0)$, and concluded that it is not a
fully satisfactory measure for HSPS, since it can, for example, be
improved by simply lowering the overall photon rate as also noted by
\cite{FATBBGZ04}. Still, from a different aspect, we have noted that
the autocorrelation at $\tau = 0$ is proportional to the variance of
the mean photon number for a source both with or without losses,
turning $g^{(2)}(0)$ into a rather good measure if related to the
mean \emph{accidental} photon number per gate $b$, which is affected 
by the rate and the gate-period.

When comparing synchronous and asynchronous HSPS, i.e.~sources with 
pulsed and CW pump lasers, regarding photon number statistics, one 
finds that both setups can in principle give either thermal or 
Poissonian original distributions. For most practical cases, a CW source 
gives a Poisson distribution, while a pulsed source gives a thermal distribution. 
By selecting temporal modes (events) from the original distributions 
by conditional gating, the photon number distribution can be further altered to show
sub-Poisson statistics, effectively decreasing both the probability of
a falsely heralded single photon, and suppressing the probability of
multiphoton events. Depending on the original photon number distribution, 
the autocorrelation shows different behaviors, giving in the ideal case 
of perfect coupling efficiencies a better result for the Poisson distribution. 
However, in a real experimental situation the two cases are practically equivalent.
In our experiment there is a probability of false heralding events 
of $52\%$, but in contrast to weak coherent pulses it is primarily of an 
experimental challenge to lower the fraction of such events by increasing 
the coupling efficiencies or the transmission factors, and of no fundamental problem.

\section{Acknowledgments}
We would like to thank Prof.~A.~Karlsson and Prof.~G.~Björk for fruitful
discussions, Dr.~A.~Fragemann, Dr.~C.~Canalias and Prof.~F.~Laurell for providing us
with the crystals, and Mr.~J.~Waldebäck for his help with electronics.
Financial support by the European Commission through the integrated
projects SECOQC (Contract No. IST-2003-506813) and 
QAP (Contract No. IST-015848), and by the Swedish 
Foundation for Strategic Research (SSF) is acknowledged.

\end{document}